\documentclass{emulateapj}

\slugcomment{}

\shorttitle{3-D structures and distances of NGC~40}
\shortauthors{Monteiro \& Falceta-Gon\c calves}

\begin{document}

\title{3-D Photoionization Structure and Distances of 
Planetary Nebulae IV. NGC 40}

\author{Hektor Monteiro\altaffilmark{1}} \affil{Departamento de
  F\'isica, Universidade Federal de Itajub\'a, Av. BPS 1303 -
  Pinheirinho, CEP 37500-903, Itajub\'a, Brazil}
\email{hektor.monteiro@gmail.com}
\author{Diego Falceta-Gon\c calves$^2$} \affil{Escola de Artes,
  Ci\^encias e Humanidades, Universidade de S\~ao Paulo, Rua Arlindo
  Bettio 1000, CEP 03828-000, S\~ao Paulo, Brazil}
\email{dfalceta@usp.br}

\begin{abstract}

  Continuing our series of papers on the 3-D structure and accurate
  distances of Planetary Nebulae (PNe), we present here the results
  obtained for the planetary nebula NGC\,40. Using data from different
  sources and wavelengths, we construct 3-D photoionization models and
  derive the physical quantitities of the ionizing source and nebular
  gas. The procedure, discussed in detail in the previous papers,
  consists of the use of 3-D photoionization codes constrained by
  observational data to derive the three-dimensional nebular
  structure, physical and chemical characteristics and ionizing star
  parameters of the objects by simultaneously fitting the integrated
  line intensities, the density map, the temperature map, and the
  observed morphologies in different emission lines. For this
  particular case we combined hydrodynamical simulations with the
  photoionization scheme in order to obtain self-consistent
  distributions of density and velocity of the nebular material.
  Combining the velocity field with the emission line cubes we also
  obtained the synthetic position-velocity plots that are compared to
  the observations. Finally, using theoretical evolutionary tracks of
  intermediate and low mass stars, we derive the mass and age of the
  central star of NGC\,40 as $(0.567 \pm 0.06)$M$_{\odot}$ and $(5810
  \pm 600)$yrs, respectively. The distance obtained from the fitting
  procedure was $(1150 \pm 120)$pc.

\end{abstract}


\section{Introduction}

Planetary nebulae are end products of the evolution of stars with
masses below $8 M_{\odot}$ and as such have great importance in many
fields in astrophysics, from basic atomic processes in solar like
stars to distant galaxies.  Although the general picture of planetary
nebulae formation is well understood \citep{K08} many questions remain
unsolved, such as the mechanism by which the stellar ejecta end up
forming the many observed morphologies.

In the past PNe have been studied with empirical methods and
one-dimensional photoionization models. This scenario has recently been
significantly changed with modern computational capabilities and
software. As we have discussed in the previous papers of this series
\citep{MSGH04,MSGGH05,SM06}, precise distances are of paramount
importance to the study of these objects. To this extent our work
provides precise, self-consistently determined distances for objects
with comprehensive, in most cases spatially resolved, observational
constraints. These objects can provide valuable calibration to
pre-existing distance scales as well as self-consistently determined
physical and chemical quantities.

One of the major limitations in our previous works is related to the
three-dimensionality of both density structure and velocity field. In
the previous papers of this series, even though we had well determined
density maps, the structures had to be defined by a combination of
parametric surfaces such as spheres and ellipsoids with assumed density
gradients to match the observations.

It is well known that hydrodynamical numerical simulations provide the
nebular 2-D and/or 3-D density distributions and the velocity
fields. Therefore, complex structures as knots, clumps or asymmetries
arise naturally from physical processes. This is particularly
important in asymmetric planetary nebulae, as the determination of
physical parameter may be compromised by simplified toy
models. Several numerical models have been proposed to explain the
nebular morphologies, such as the two-wind interaction model
\citep{IBF92}, jets \citep{AS08}, magnetic fields \citep{GLF05}, and
even binarity, which accounts for anisotropic distribution of the AGB
wind, or for the formation of bipolar jets.

Actual density distributions obtained from these simulations are, in
general, directly compared to observed maps. However, it is well known
that the gas density distribution - or even the column density
projection along the line of sight (LOS) - may differ when compared to
the observed maps. The difference may arise because the observations
depend on the projected fluxes of certain spectral lines, which in
turn depends on the three-dimensional ionization structure. In this
sense, only combined calculations of both hydrodynamics and
photoionization processes can provide realistic emission maps.

In this work we present a self-consistent model fit for NGC~40 using
this new approach, employing both hydrodynamical and
photoionization schemes. In the next section we present the current
knowledge about NGC~40 and the observational data used in our
analysis. In Section 3 we present the numerical scheme and the
hydrodynamical simulations for this object. The photoionization model
is described and the results shown in Section 4. Following we discuss
the kinematics of the nebular material combining the synthetic
emissivities and the simulated data in Section 5, followed by the
discussions and conclusions.

\section{NGC~40}

NGC~40 is a well studied object, classified as a low excitation PN with
a WC8 type central star. Several imaging studies revealed a bright
(slightly elliptical) core, a large halo and filamentary structures
\citep{CJA87,B87,BGFJ92,MLBM96}.  \cite{CSPT83} presented optical and
IUE spectra for the bright region of the nebula. Their results
indicated that the abundances were typical and did not display large
variations, expected for a WC8 central star. The authors argue that
the CIV 1549 emission from the nebular envelope is too strong to have
been produced by normal thermal processes sugesting that it is a
consequence of processes related to the wind of the central star.

\cite{MLBM96} obtained narrow band CCD images as well as long slit
echelle spectra and showed that the outer halo is moving with 31km
s$^{-1}$, with a turbulent motion of ~7km s$^{-1}$.  The electron
temperature for that region was calculated to be $T_e=(7400 \pm
160)$K. The inner halo line profiles are split by about 50 km s$^{-1}$
which is consistent with the splitting seen in the bright core. The
main outer clumps seen in narrow band images are shown to be
kinematically associated with barrel shaped core, thus not being the
result of a true jet. The authors also find a dynamical age of $\sim
4000$yr from the expansion of the core. \cite{KDR98} and the dynamical
ages $3500 \pm 500$yr for NGC 40.

In \cite{SCBTZ00} the authors present detailed tomography of NGC~40,
obtained with echelle spectroscopy in many slit positions. By assuming
a simple, direct position-velocity correlation they reconstruct the
spatial distribution of $H^+$, $O^{++}$ and $N^+$. The results show a
complex density structure with densities as high as 3000$cm^{-3}$. The
high densities found are in agreement with results obtained by
\cite{LGMR10}(LGMR10 hereafter) who obtained values around
2500$cm^{-3}$, although the errors are large due to the weak lines
used. The expansion velocities determined by \cite{SCBTZ00} are around
25km s$^{-1}$. The authors also point out that abundance gradients are
likely to be present in the nebula, which is also consistent with the
abundance maps of LGMR10, where the authors present the first
spatially resolved spectroscopic mapping of NGC~40.

In the infrared bands, \cite{HWT01} reported the discovery of
$21\micron$ and $30\micron$ emission features in the spectrum obtained
from the ISO observatory. Based on the presence of the $21\micron$
feature, they argue that the bulk of the dust in the nebula has been
produced during a carbon-rich phase before the atmospheres of these
stars became hydrogen poor. The authors also present an estimation of
the dynamical age of the object of $\sim 5000 yr$. However,
\cite{CBL02} ascribed the $21\micron$ feature to noise.

When considering the distance determinations, NGC~40 follows more or
less the tendency of large interval that most PNe determinations
have. From the Acker catalog \citep{AO92} we find $<d>=1019 \pm 357$pc, 
ranging from 620 to 2070 pc.

\subsection{Observational data}

The observational data used to constrain our model parameters in the
present work was taken from different sources of the literature. We
use spatially resolved emission line maps as our main constraint in
the ionization structure and spatial density and temperature
distribution. The spatially resolved data comes from the work of
LGMR10 described previously. We reffer the reader to that work and
references therein for greater detailed discussion.

For the absolute $H\beta$ flux we use the value of \cite{CSC83}. The
main reason is that the data obtained in LGMR10 was calibrated using
only one star and the sky was not considered to be in photometric
conditions. In this sense, the fluxes obtained from the LGMR10 data set
are good for relative intensities only.  We also use the integrated
fluxes of \cite{LLLB04} (hereafter LLLB04) to help constrain the model
integrated fluxes. Other works mentioned in the introduction were also
used but with less weight in the final comparisons.

In the infrared band we used data values for NGC~40 obtained from the
2MASS survey, the IRAS point source catalog and also spectrum from the
ISO archives\footnote[1]{ISO data tag:
  ADS/IRSA.Atlas\#2010/1007/114440\_6120}. We also use fluxes at 9 and
18 $\mu$m obtained with the Infrared Camera (IRC), and at 65, 90, 140
and 160 $\mu$m using the far-Infrared Surveyor (FIS) derived from the
AKARI All-Sky Survey made available in \cite{PM11}.

In order to model the kinematics of NGC~40 we used as constraints the
line velocities obtained by \cite{SCBTZ00}, which obtained spatially
resolved, long-slit echellograms of the nebula at $U$, $B$ and $V$
filters and for different position angles. This data is used for the
construction of synthetic tomographic maps of the nebula that are
directly compared to the observations.

\section{Hydrodynamical Simulations of NGC~40}

In order to obtain more realistic density and velocity distributions
for the photoionization models, we performed a number of 2.5D
hydrodynamical simulations of planetary nebula ejecta.

We have employed the grid code Godunov-MHD, which solves the gas
hydrodynamical equations in conservative form. The effects of
radiative cooling are calculated implicitly after each time step (see
Falceta-Gon\c calves, Lazarian \& Houde 2010, Falceta-Gon\c calves et
al. 2010a,b).  The high-order shock-capturing scheme is based on an
essentially non-oscillatory spatial reconstruction and Runge-Kutta
time integration. The discontinuities that arise in simulations of
supersonic expanding shells are better solved using the HLLC Riemann
solver \citep{londrillo00}.

\begin{figure}[!ht]
\centering
\includegraphics[width=\columnwidth]{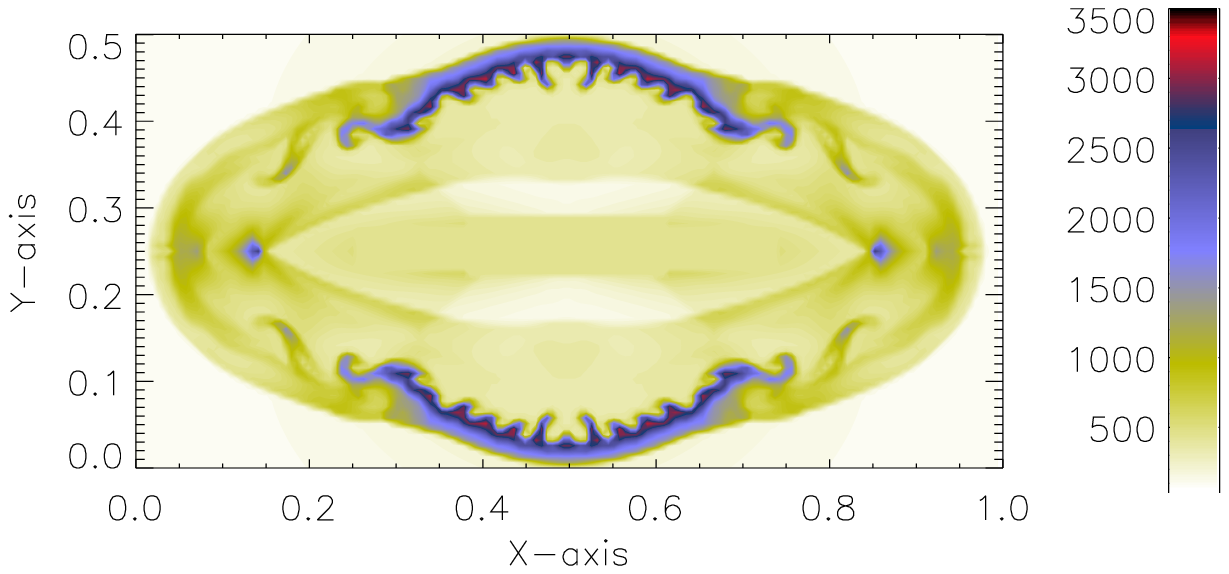}
\includegraphics[width=\columnwidth]{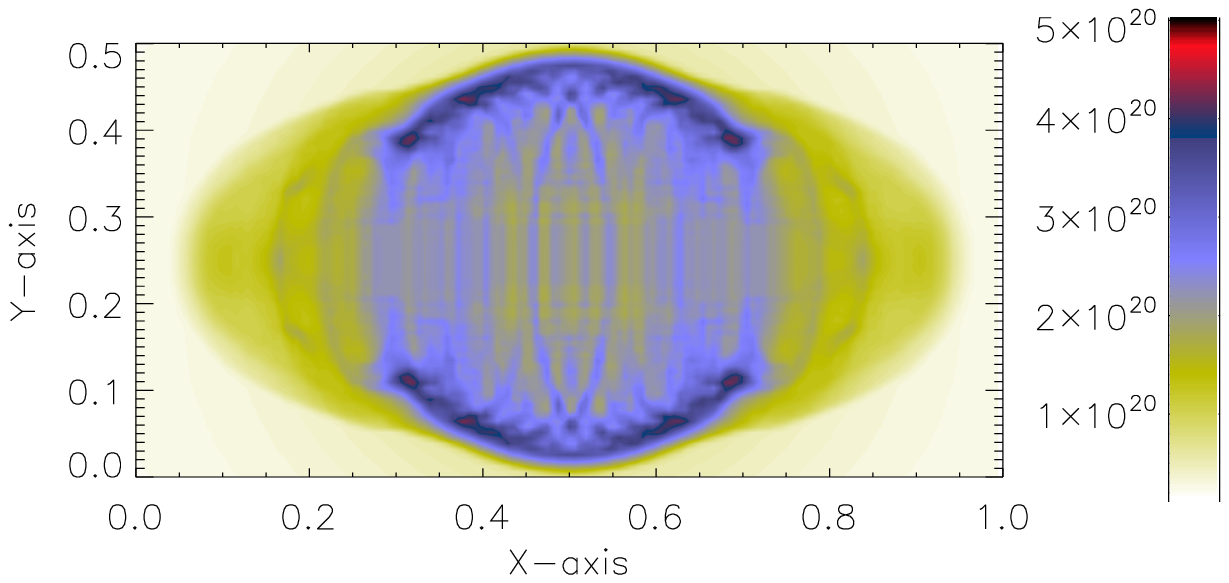}
\caption{ Density (up) and column density (bottom) maps for an
  inclination of $20^{\circ}$ with respect to the plane of sky
  . \label{dens} }\end{figure}

\subsection{Setup for PN ejection}

As initial condition for this problem we use the basic model of two
interacting winds \citep{IBF92}, which represents the
ejection of the tip-AGB stellar envelope over a preset density
distribution given by the previous wind of the red giant star. To
account for the giant-phase stellar wind density profile we set the
ambient gas density as \citep{IPB89}:

\begin{equation}
\rho_{\rm amb}=\frac{\rho_0}{A(\theta)}\left( \frac{r_0}{r} \right) ^2 ,
\end{equation}

\noindent
where

\begin{equation}
A(\theta) = 1-\alpha \left( \frac{e^{\beta cos 2 \theta - \beta}-1}{e^{2 \beta}-1} 
\right).
\end{equation}

The parameter $\alpha$ is related to the density ratio at the polar
and equatorial directions, while $\beta$ the steepness of the density
profile with the latitude. The initial temperature of the environment
is defined as $100$K.

As mentioned previously, in the past two decades it has become clear
that the two-wind interaction model is unable to reproduce the
morphologies of most PNe. Many alternative models have been proposed
including other dynamical mechanisms for the ejection and shaping of
the PNe, such as jets (Sahai \& Trauger 1998, Lee \& Sahai 2003, Lee
\& Sahai 2004, Akashi \& Soker 2008, Lee, Hsu \& Sahai 2009) and
magnetic fields (Garcia-Segura 1997, Garcia-Segura et al. 1999,
). From the theoretical point of view, if the magnetic pressure is of
order of the kinetic and thermal energy densities we should expect a
larger degree of collimation of the lobes. A similar effect would be
expected in a “heavy jet” scenario. In a situation with any of these
two mechanisms the resulting $\alpha$ and $\beta$ parameters would be
slightly different. In this work, however, we decided to use the
simpler two-wind interaction model in order to reduce the number of
free parameters in modeling the dynamical properties of this specific
object.

We performed 2.5D simulations for different sets of $\alpha$ and
$\beta$ parameters in order to obtain the most similar projected
density distribution to the observed morphology for NGC~40. We run
models varying $\alpha$ from 0.0 to 0.9, in steps of 0.1, and $\beta$
from 0.0 to 6.0, in steps of 0.5. Due to computational limitations it
was impossible to run the models in 3-dimensions. All models were
calculated with a fixed grid of $512 \times 256$ resolution. The cell
size is defined as $2.2 \times 10^{15}$cm and the simulation box is
$1.15 \times 10^{18}$cm wide.  The simulations were run up to a
dynamical time of $10000$yrs. The structure adopted for the nebulae is
constrained by the observed density map and the observed projected
morphology. The best model for this particular object was obtained for
$\alpha = 0.3$ and $\beta=5.0$, as described below.

\subsection{Hydrodynamical results}

The initial distribution of density is determined by the anisotropic
wind from the AGB-star. At the center of the cube, the ejection of the
second and faster wind compresses the environment gas resulting in the
nebular structure. The initial anisotropy of the pre-nebular wind and
of the post-ejection wind results in a roundish structure with two
lobes. In Figure 1 (up) we show the density distribution obtained at
$t=5900$yr.  At this time the size of the structure is $\sim 61500$ AU
wide. The nebular peak density is $\sim 2800$cm$^{-3}$ at the roundish
structure. Here, the visible filaments are formed by Rayleigh-Taylor
instability that arise as the lighter and warmer plasma expands onto
the cold and dense structure.

At both opposing sides along the major axis of the nebula, the round
structure is opened forming two lobes. As the internal wind material
propagates outwards it reaches the inner shock. Since the expanding
gas is not isotropic, the streamlines cross the shock surface
obliquely resulting in a deviation of the flow. The material flowing
along the nebular shell then converges at the apex of the lobe, in the
so-called ``shock focused inertial confinement”, resulting in a dense
structure, as seen in Figure 1, and eventually in a jet (Frank, Balick
\& Livio 1996). At the stage of the nebula in this simulation there is
no jet formed. The lobes present numerical densities of $\sim
1000$cm$^{-3}$, except for the two opposing knots - also formed by
RT-instability - at the edges of the major axis of the structure,
which are $50 \%$ denser. The ejecta average velocity is $\sim 25 $km
s$^{-1}$.

Besides the main nebula seen with densities $> 1000$cm$^{-3}$, a
diffuse gas with densities $\sim 500$cm$^{-3}$ surrounds the whole
structure. The inner surface is determined by the reverse shock of the
main nebula with the expanding hot wind of the central source. The
white dwarf wind blows the hot and low density ($< 200$cm$^{-3}$) gas
which is, in general, observed in X-rays.

Despite of the fact that the simulation is performed in 2.5D, the
structure may be considered of cylindrical symmetry. Therefore, it is
possible to rotate the physical variables around the nebular major
axis of symmetry in order to obtain a three dimensional
distribution. We must point out here that when this approximation is
used the small scale structures, such as those generated by the RT
instabilities mentioned above, will appear in the column density maps
as circles/rings instead of knots, as would be expected from 3D
simulations. For the purposes or the present work this limitation does
not influence our results. With the cube of density it is possible to
integrate it and obtain the column density for several LOS's. In
Figure 1 (bottom), we illustrate the projected column density for an
inclination angle of $20^{\circ}$ for the major axis with respect to
the line of sight. The column density map shows similar features as
seen in the density map, except for the missing two opposing knots
located at the end of the major axis of the nebula. The reason for
that is the small size of these clumps.  Interestingly, two opposing
knots at the major axis are visible in the observed maps of NGC~40. As
we show further in this paper, the synthetic maps obtained after
computing the photoionization reveal the existence of the knots for
low ionization lines, such as SII and NII.

\section{Photoionization Models for NGC 40}

In the present work we used the photoionization code Mocassin~3D
(version 2.02.54) described in full detail in \cite{EBSL03}.

The procedure to study NGC~40 is the same adopted in previous papers
of this series, i.e. we provide a density distribution of the nebular
plasma and run the Monte-Carlo simulation using the Mocassin~3D code
for the radiative transfer, assuming a given luminosity ($L_\nu$) from
the central source. The whole process is iterated until reasonable
agreement with observational constraints is achieved. The main
improvement of this work with respect to the previous ones is the use
of a self-consistent distribution of density from the numerical
hydrodynamical simulations. In the following subsections we discuss
the input parameters used.

\subsection{The ionizing source}

The choice of the ionizing source in photoionization models is far
from being the trivial choice of a blackbody curve of a given
temperature. Recent developments in the modeling of stellar
atmospheres have provided more complex stellar spectra that can be
used with physical and chemical parameters, such as $log(g)$ and
abundances of the elements which may be important.

In the field of PNe central stars (CS) the standard choice of models
is the grid provided by T. Rauch \footnote[2]{available at
  http://astro.uni-tuebingen.de/~rauch/} where the user can find NLTE
stellar model atmosphere fluxes which cover the parameter range of PNe
ionizing sources: $T_{eff}$ = 50 - 190 kK, $log(g)$ = 5 - 9
($cm/sec^2$) and distinct abundances.

\begin{figure}[!h]
\includegraphics[width=\columnwidth]{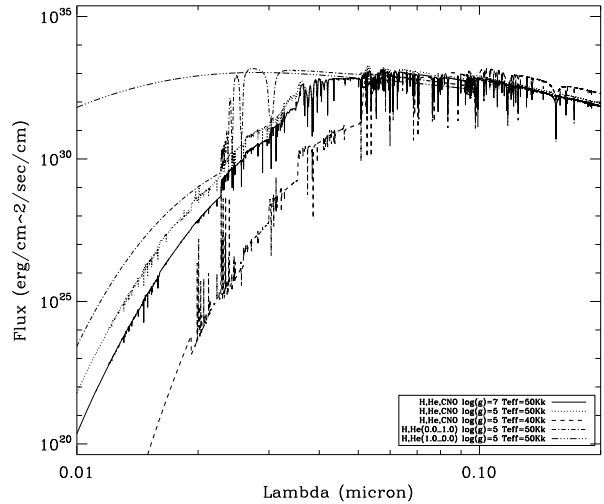}
\caption{ Central star ionizing spectra studied in this work. }
\end{figure}  

\begin{table*}[ht]
\small
\begin{center}

\caption{Observed and model line fluxes and model central star 
parameters for NGC\,40.\label{mod-res}}
\begin{tabular}{llllllll}
\hline
   &   &   &  &  &  &  & Observed$^a$ \\
\hline\hline 
H:He:CNO         &  1:0:0 &  0:1:0 &  0:1:0 &  0:0.33:0.67  &  0:0.33:0.67 &0:0.33:0.67 &                         \\
Teff                 &  50kK  &  50kK  &  50kK  &  40kK  &  50kK  &  50kK  &                            \\
log(g)               &  5     &    5   &  7     &  5     &  5     &  7     &                            \\
H$\beta$ flux$^*$    &  0.032 &  0.034 &  0.031 &  0.033 &  0.037 &  0.036 & 0.038                       \\
He~{\sc ii}4686      &  0.157 &  0.000 &  0.000 &  0.000 &  0.000 &  0.000 & 0.000                       \\
$[$N~{\sc ii}$]$5755 &  0.051 &  0.041 &  0.064 &  0.03  &  0.04  &  0.05  & 0.03                        \\
$[$N~{\sc ii}$]$6548 &  0.774 &  0.705 &  0.852 &  0.98  &  0.84  &  0.90  & 0.85                         \\
$[$N~{\sc ii}$]$6584 &  2.365 &  2.154 &  2.602 &  2.99  &  2.56  &  2.75  & 2.54                         \\
$[$O~{\sc ii}$]$3726 &  0.078 &  0.066 &  0.240 &  1.110 &  0.881 &  2.381 & 2.571$^b$                     \\
$[$O~{\sc ii}$]$3729 &  0.052 &  0.043 &  0.149 &  0.717 &  0.534 &  1.442 & 1.904$^b$    
                 \\
$[$O~{\sc iii}$]$4363&  0.006 &  0.005 &  0.007 &  0.000 &  0.004 &  0.004 & 0.005                         \\
$[$O~{\sc iii}$]$4959&  0.193 &  0.170 &  0.189 &  0.000 &  0.172 &  0.196 & 0.193                         \\
$[$O~{\sc iii}$]$5008&  0.576 &  0.509 &  0.563 &  0.000 &  0.514 &  0.585 & 0.589                          \\
                                                                
\hline 
                                
Line Diagnostics        &    &     &      &  &  &  &  \\               
Ne:                     &    &     &      &  &  &  &  \\                                
$[$S~{\sc ii}$]$6731/6717         & 1.166 &   1.206 &   1.242 &   1.236  & 1.275 & 1.261 & 1.308\\
$[$O~{\sc ii}$]$3729/3726         & 0.667 &   0.650 &   0.621 &   0.640  & 0.606 & 0.605 & 0.739 \\
Te:                     &    &     &      &    &  &  &                \\
$[$N~{\sc ii}$]$(6584+6548)/5755  & 61.826 &  69.695 &  54.138 &  132.8 & 82.9 & 80.7 & 74.4 \\
$[$O~{\sc ii}$]$3726/7320         & 19.005 &  20.575 &  16.052 &  33.8  & 21.7 & 21.3 & 36.7\\
$[$O~{\sc iii}$]$(4959+5007)/4363 & 126.95 &  152.190 &  103.041&  - & 171.6 & 195.3 & 156.4 \\

\hline
$^b$ relative fluxes from \cite{LGMR10}\\
$^b$ relative fluxes from \cite{LLLB04}\\
$^*$$1 \times 10^{36} erg cm^{-2} s^{-1}$
\end{tabular}
\end{center}
\end{table*}

We have explored a set of CS models to obtain a best match to the
observed PN line intensities in NGC~40. The different spectra used are
shown in Figure 2. The photoionization code was run with the density
distribution obtained from the hydrodynamical simulation, shown in
Figure 1 (transformed in 3-D). The plots differ from $log(g)$,
$T_{eff}$ and chemical abundances, which are described in Table 1. We
found best fitting models for each CS star studied giving more
emphasis on reproducing the total $H\beta$ flux first and then other
emission lines. The final step consisted of fine tuning the abundances of
elements that are more sensitive to the $T_{eff}$ such as oxygen, in
an attempt to reproduce the observed line fluxes.

As expected, the spectra seem very similar at the region that
contributes most for the ionization of H, making it hard to select an
ionizing source based only on the total $H\beta$ flux. To pin point
the best spectrum we must analyze other emission lines produced at
different regions of the nebula, and by ions sensitive to the ionizing
source temperature. We point out that the HeII 4686 emission line is
one of the best to constrain $T_{eff}$, however it is too weak in
NGC~40 and could not be used.  The most relevant differences appear in
the oxygen lines as can be seen in Table 1, where the last column
shows the observed values obtained from LLLB04 and LGMR10.

From the data presented in Table 1 we infer that the best ionizing
spectrum was the one with [H:He:CNO] 0:0.33:0.67, which are typical
values for a PG 1159 type star (hydrogen-deficient post-asymptotic
giant branch stars believed to be descendents of [WC] type stars). For
a very detailed review of PG 1159 stars see \cite{WH06}. Obviously,
due to computational limitations, the grid search for the ionizing
source was not exhaustive. However we explored the most likely
alternatives given the current knowledge of PNe ionizing sources and
observation data for NGC~40.

\subsection{Dust}

Apart from the main emission lines, NGC~40 also presents large
continuum fluxes that are related to the dust component, as observed
in infrared bands. Therefore, we have also included dust in the
modeling, taking advantage of the full potential of the Mocassin 3-D
code. As discussed previously, some authors have already dealt with
the infrared data and dust content of this object. The main goal in
the present work is to establish reasonable intervals for the amount
of dust present as well as some information on its properties, rather
than obtaining an excellent fit for the infrared observations.

To constrain the model dust parameters we have gathered the spectra
obtained by ISO as well as photometry in the IRAS catalog. The ISO
spectra was obtained from the NASA/IPAC Infrared Science Archive. We
also use fluxes at 9, 18, 65, 90, 140 and 160 $\mu$m using the data
made available in \cite{PM11}.

The input parameters for the dust obtained from the fitting procedure
were, assuming the dust to be perfectly mixed with the gas, a mass
dust to gas ratio of 0.0015 with a typical MRN size distribution and
composition of 50\% silicates, 40\% grafites and 10\% PAH grains. The
data tables for the dust parameters are those available in the
Mocassin dust data directory and we refer the reader to \cite{EBS05}
for details.

\section{Photoionization model results}

In this section we present the main results from the final fitted
model for NGC~40.

\subsection{Synthetic observational maps of NGC~40}

\begin{figure*}
\includegraphics[scale=1.0]{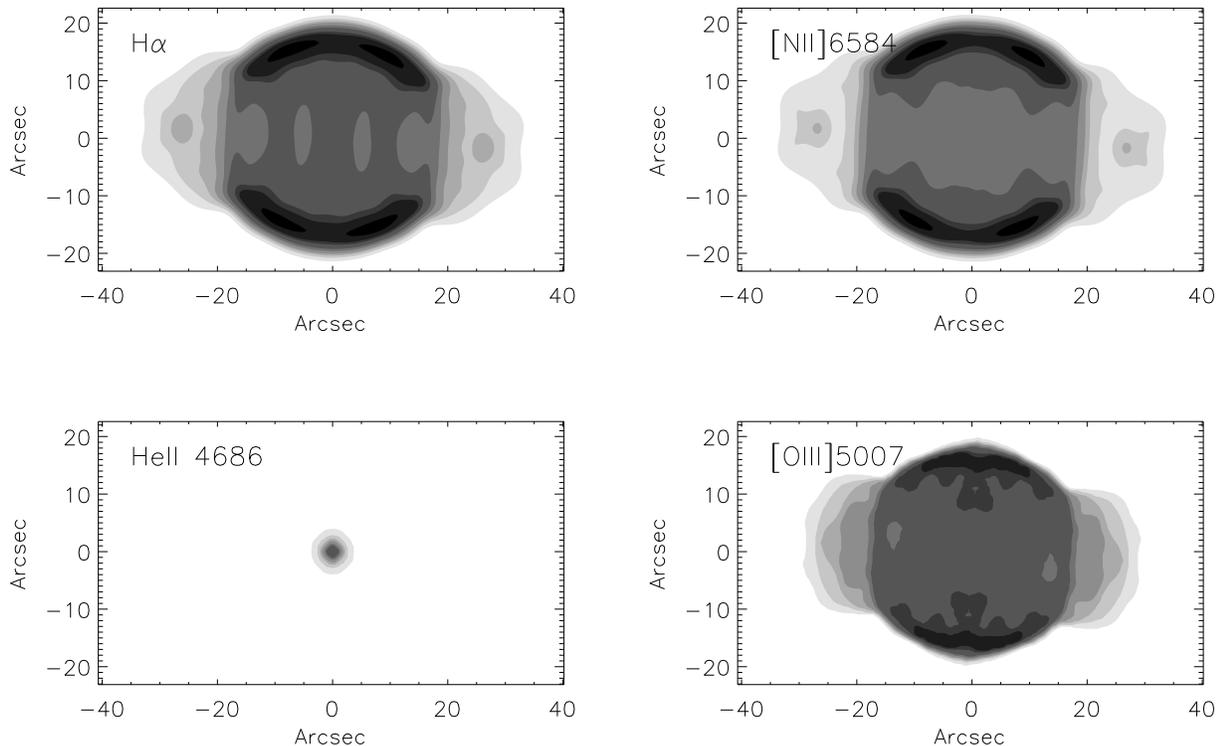}

\caption{Synthetic images obtained from the projection of the data cubes of
  emissivities calculated by the photoionization code for 4 of the most
  important emission lines. The nebula has been rotated with an angle 
of 20$^{\circ}$ with respect to the plane of sky. 
\label{modims6781} }

\end{figure*}

Once the emissivity at each wavelength is obtained, from the best fit
photoionization model, it is possible to construct the synthetic
observational maps of NGC~40. In order to accomplish this we
integrated the emissivity cube along a given line of sight. In Figure
3 we show the synthetic maps obtained for the $H\alpha$, [NII]6584,
HeII and [OIII]5007 spectral lines assuming an inclination of
20$^{\circ}$ for the nebular major axis with respect to the plane of
sky.

\begin{figure}[!h]
\includegraphics[width=\columnwidth]{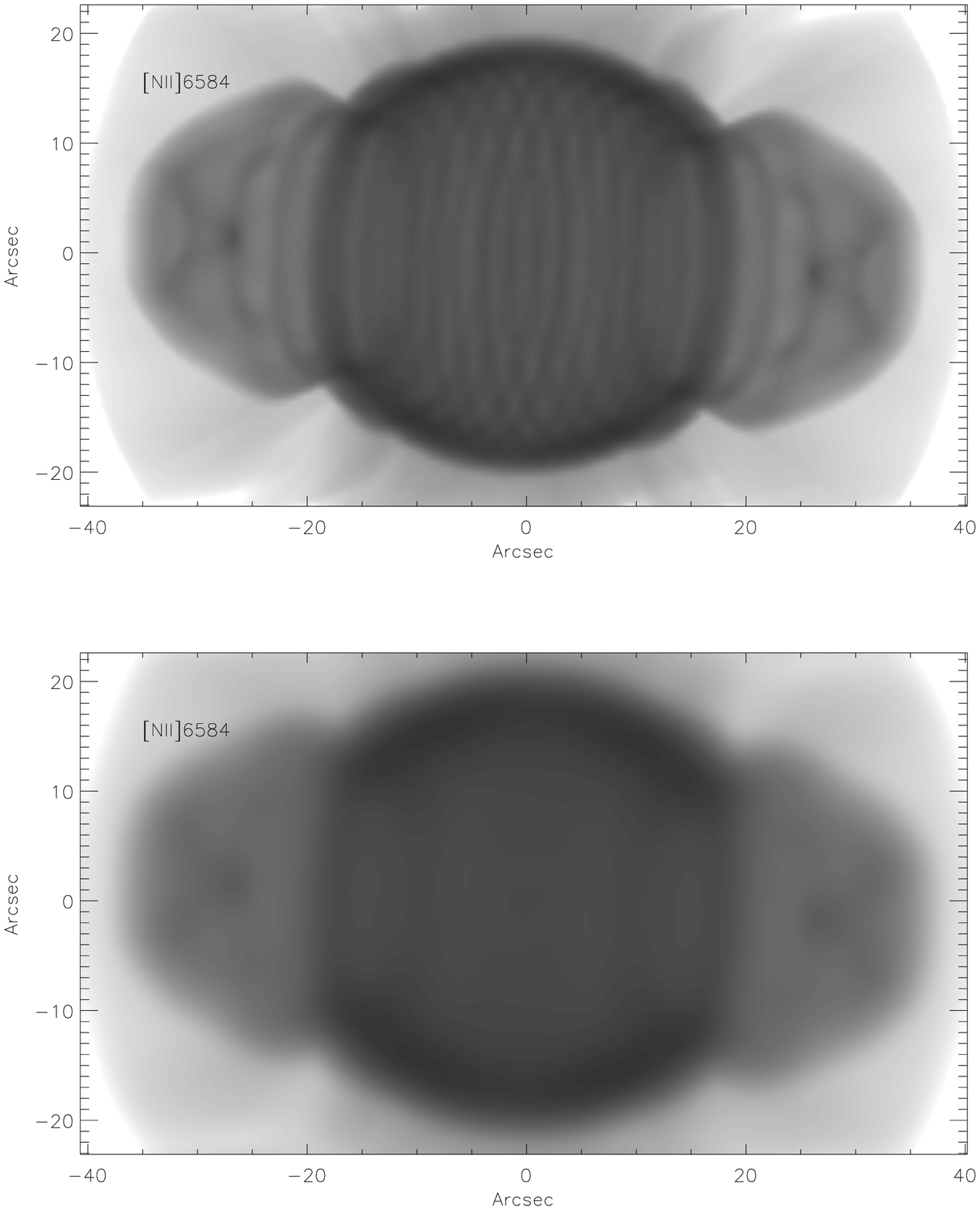}
\caption{ Model [NII]6584 images obtained from the best fit
  model. Upper image shows the raw model cube projected with
  orientation as described in the text and lower image shows the same
  projection but convolved with a 3 arcsec PSF to match the resolution
  of LGMR10 maps.}
\end{figure}

These synthetic maps cannot be directly compared to the observations
due to the different spatial field resolution. The numerical
simulations presented here present finer resolution when compared to
the observed maps of LGMR10. In this sense, in order to match their
spatial resolution we convolved the synthetic maps with a 3 arcsec
PSF. In Figure 4 we show the original synthetic map for the emission
line [NII]6584 (top) and the convolved map for the same line
(bottom). Despite the coarser spatial resolution, the knots seen at
the ends of the major axis of symmetry are still visible and the
general morphology of the nebula is kept. At smaller scales, the
inhomogeneities created by the Rayleigh-Taylor instabilities are
smoothed.

\begin{figure}[!h]
\includegraphics[width=\columnwidth]{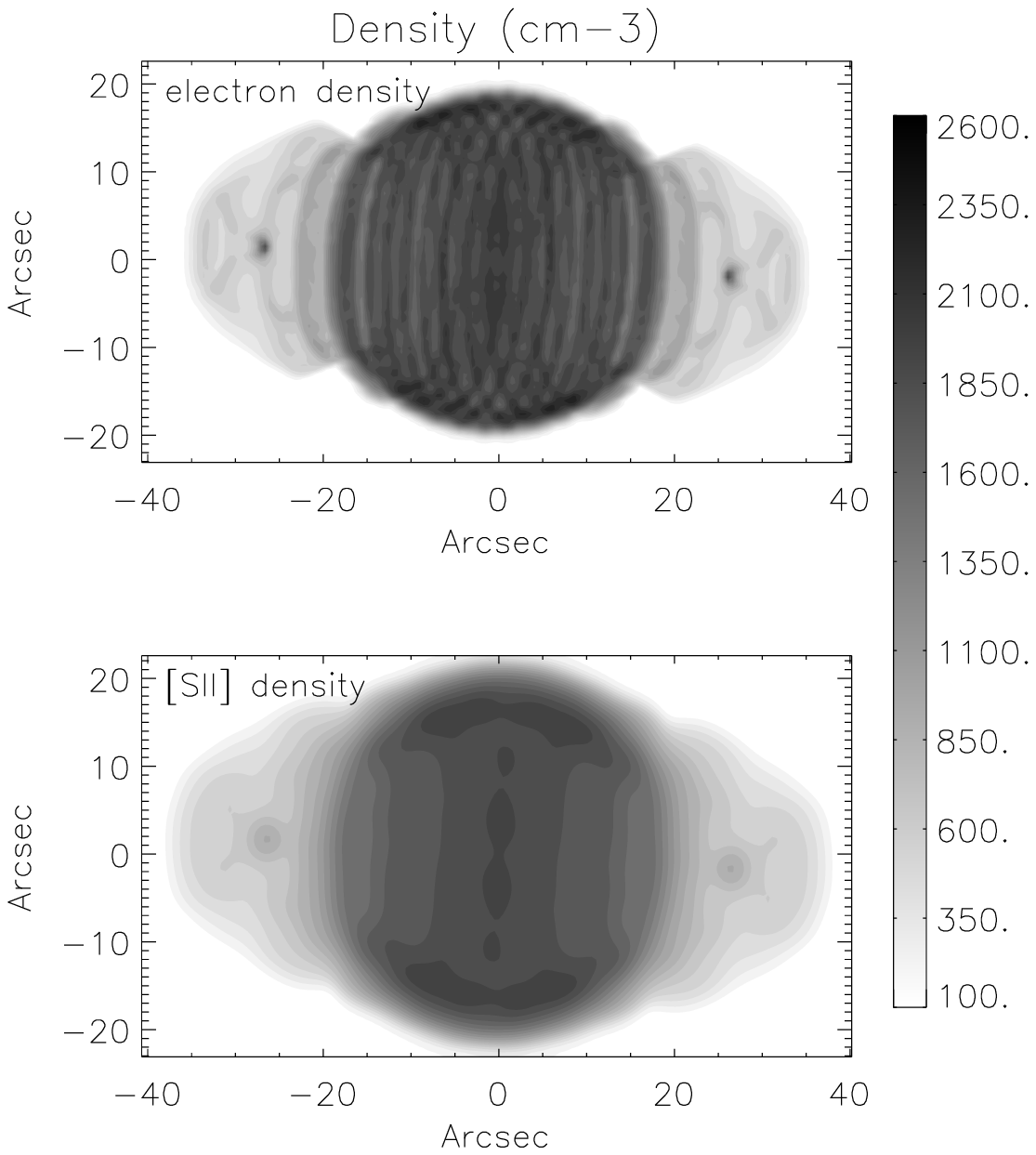}
\caption{Electron density maps obtained from the photoionization
  model. Upper panel shows projected electron density cube, weighted
  by the $[SII]6717$ emissivity cube. Lower panel shows electron
  density obtained from the usual diagnostic ratio using the convolved
  emission line maps.}
\end{figure}  

In Figure 5 we present two density maps obtained from the model. The
first map in the upper panel was obtained from the projection of the
electron density data cube weighted by the $[SII]6717$ emissivity
cube. The lower panel shows the density map obtained from the usual
ratio of the convolved $[SII]$ emission line maps. The two show good
agreement despite the different methods used in the calculation.

In Figure 6 we present two electron temperature maps obtained from the
model. The first map in the upper panel was obtained from the
projection of the electron temperature data cube weighted by the
$[OIII]4363$ emissivity cube. The lower panel shows the electron
temperature map obtained from the usual diagnostic ratio
$[OIII]5007+4959/[OIII]4363$, again using the convolved emission line
maps. Again, apart from the difference in resolution we see good
agreement in the values obtained.

\begin{figure}[!h]
\includegraphics[width=\columnwidth]{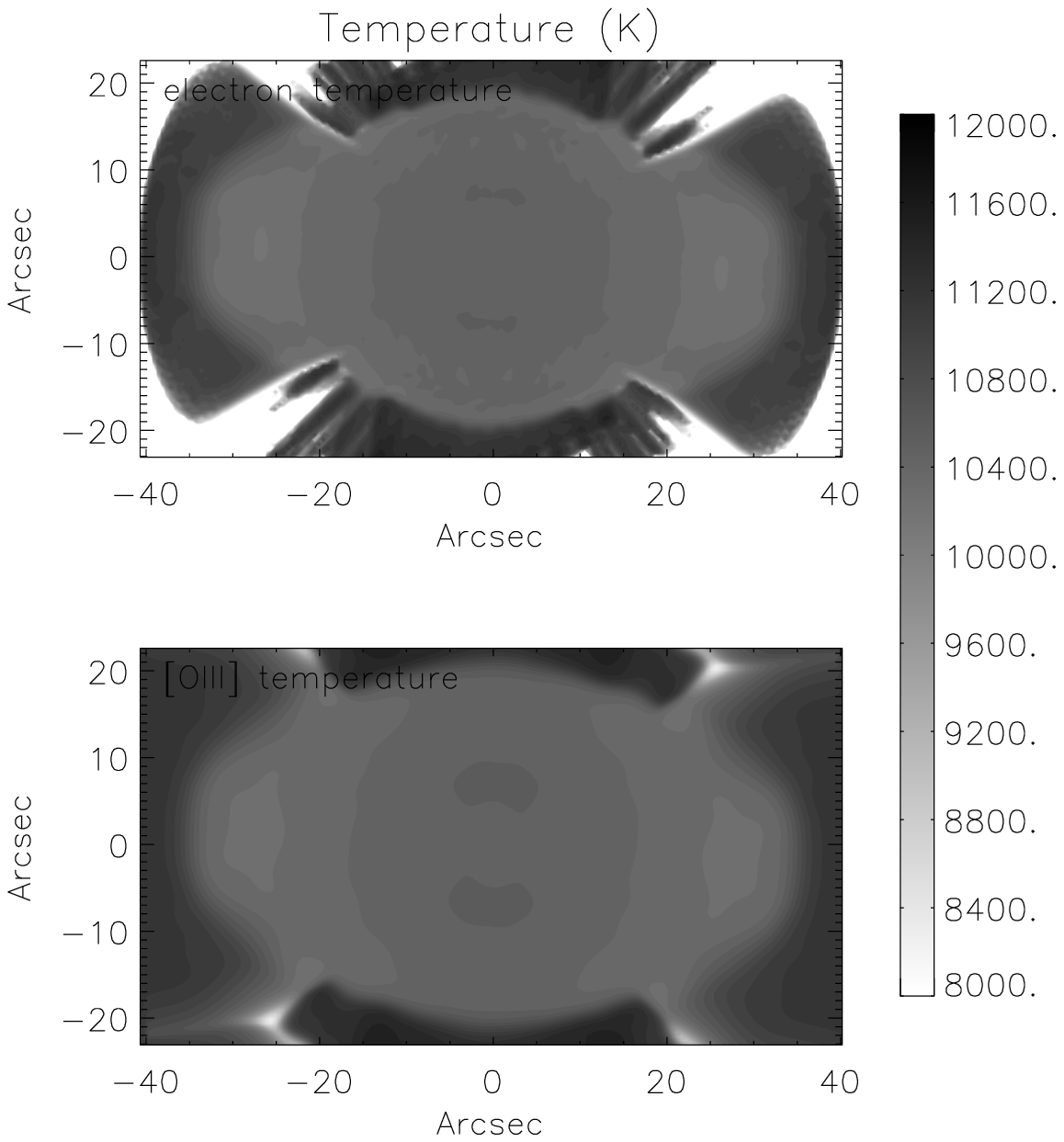}
\caption{Electron temperature maps obtained from the photoionization
  model. Upper panel shows projected electron temperature cube,
  weighted by the $[OIII]4363$ emissivity cube. Lower panel shows
  electron temperature obtained from the usual diagnostic ratio using
  the convolved emission line maps. }
\end{figure}  

In Figure 7 we present the observed $H\alpha$ narrow band emission map
superimposed by the contours of the convolved synthetic map from the
best fit model (taking into account the $\sim$15\% [NII] contamination
in the narrow band filter). The agreement between both, the size and
morphology of the true nebular emission, and the presented model is
clear.

The total line intensities of the best fit model are given in
Table\,\ref{mod-res}, as well as the fitted abundances and ionizing
source parameters. To estimate a final relative uncertainty to be used
in the fitted model parameters we calculated the relativer errors for
all lines in Table\,\ref{mod-res}, using the average of the observed
values when both LLLB04 and LGMR10 are present. The final relative
uncertainty can be obtained by taking a weighted average of all
relative errors, using line intensities as the weights. The value
obtained is 10\% which we adopt as the best estimate of the relative
$1\sigma$ uncertainty of our final fitted parameters.

The model fitting procedure which uses, among other constraints, the
model image size fitted to the observed one for $H\alpha$, as well as
the absolute $H\beta$ flux, gives a final distance of 1150$\pm$120\,pc
for NGC~40.
   
\begin{table}[!ht]
\begin{center}
\caption{Observed and model line fluxes and model central star 
parameters for NGC\,40.\label{mod-res}}
\begin{tabular}{llll}\hline  &   {LLLB04}&   {LGMR10} &   {ModPG}  \\
\hline\hline 
\\
{$T_*$ (kK)} & {} & {}& {50}\\
{$L_*/L_{\odot}$} & {} & {}& {1736}\\
{$log(g)$} & {} & {}& {7}\\
{Density ($cm^{-3}$)} & {1750} & {100-2700}& {100-2600}                       \\
{He/H}      & {$1.2\times 10^{-1}$}& {$1.2\times 10^{-1}$}& {$0.9\times 10^{-1}$}                       \\
{C/H}       & {$6.9\times 10^{-4}$}& {-}                 & {$6.5\times 10^{-4}$}                        \\
{N/H}       & {$8.5\times 10^{-5}$}& {-}                 & {$6.8\times 10^{-5}$}                        \\
{O/H}       & {$4.9\times 10^{-4}$}& {$4.1\times 10^{-5}$}& {$1.9\times 10^{-4}$}                        \\
{S/H}       & {$2.6\times 10^{-6}$}& {-}                 & {$6.8\times 10^{-6}$}                        \\
{Cl/H}      & {$8.1\times 10^{-8}$}& {-}                 & {$7.3\times 10^{-8}$}                        \\
\hfill \\
{log($H\beta$)}     &  {-9.62}& {-} &  {-9.61}           \\

[O~{\sc ii}]~3726      & {2.57} & - & 2.61   \\

[O~{\sc ii}]~3728      & {1.90} & - & 1.52   \\
H10~3797               & {0.05} & 0.04 & 0.05   \\
H9~3835                & {0.08} & 0.07 & 0.07   \\
H8$+$He~{\sc I}~3888   & {0.18} & 0.17 & 0.22   \\
H$\delta$~4101         & {0.28} & 0.22 & 0.26   \\
H$\gamma$~4340         & {0.50} & 0.44 & 0.47   \\
$[$O~{\sc iii}$]$~4363 & {0.003} & 0.006& 0.004   \\
He~{\sc i}~4471        & {0.03} & 0.02 & 0.05  \\
He~{\sc ii}~4686       & {0.00} & 0.01 & 0.00   \\

He~{\sc i}~4921        & {0.01} & 0.01 & 0.01   \\
$[$O~{\sc iii}$]$~4959 & {0.14} & 0.19&  0.19  \\
$[$O~{\sc iii}$]$~5007 & {-}    & 0.59 & 0.57   \\
$[$N~{\sc i}$]$~5198   & {0.01} & 0.00 & 0.00   \\
$[$Cl~{\sc iii}$]$~5517 & {0.004} & 0.004 & 0.005   \\
$[$Cl~{\sc iii}$]$~5537 & {0.004} & 0.005& 0.005   \\
$[$O~{\sc i}$]$~5577    & {-} & 0.006 & 0.00   \\
$[$N~{\sc ii}$]$~5755   & {0.03} & 0.03 & 0.04   \\
He~{\sc i}~5876         & {0.09} & 0.11 & 0.13   \\
$[$O~{\sc i}$]$~6300    & {0.03} & 0.03 & 0.02   \\
$[$O~{\sc i}$]$~6363    & {0.01} & 0.01 & 0.01   \\
$[$N~{\sc ii}$]$~6548   & {0.87} & 0.85 & 0.83   \\
H$\alpha$~6563          & {2.99} & 2.97 & 2.85   \\
$[$N~{\sc ii}$]$~6584   & {2.67} & 2.54 & 2.94   \\
He~{\sc i}~6678         & {0.03} & 0.02 & 0.03   \\
$[$S~{\sc ii}$]$~6717   & {0.12} & 0.13 & 0.14   \\
$[$S~{\sc ii}$]$~6731   & {0.16} & 0.19 & 0.18   \\
$[$C~{\sc iv}$]$~1550   & {0.26} & - & 0.00   \\
C~{\sc ii}~2841         & {0.04} & - & 0.007   \\
$[$C~{\sc ii}$]$~158$~\micron$   & {0.038} & - & 0.036   \\

\hline
\end{tabular}
\end{center}

\end{table}

\begin{figure}[!h]
\includegraphics[scale=0.6]{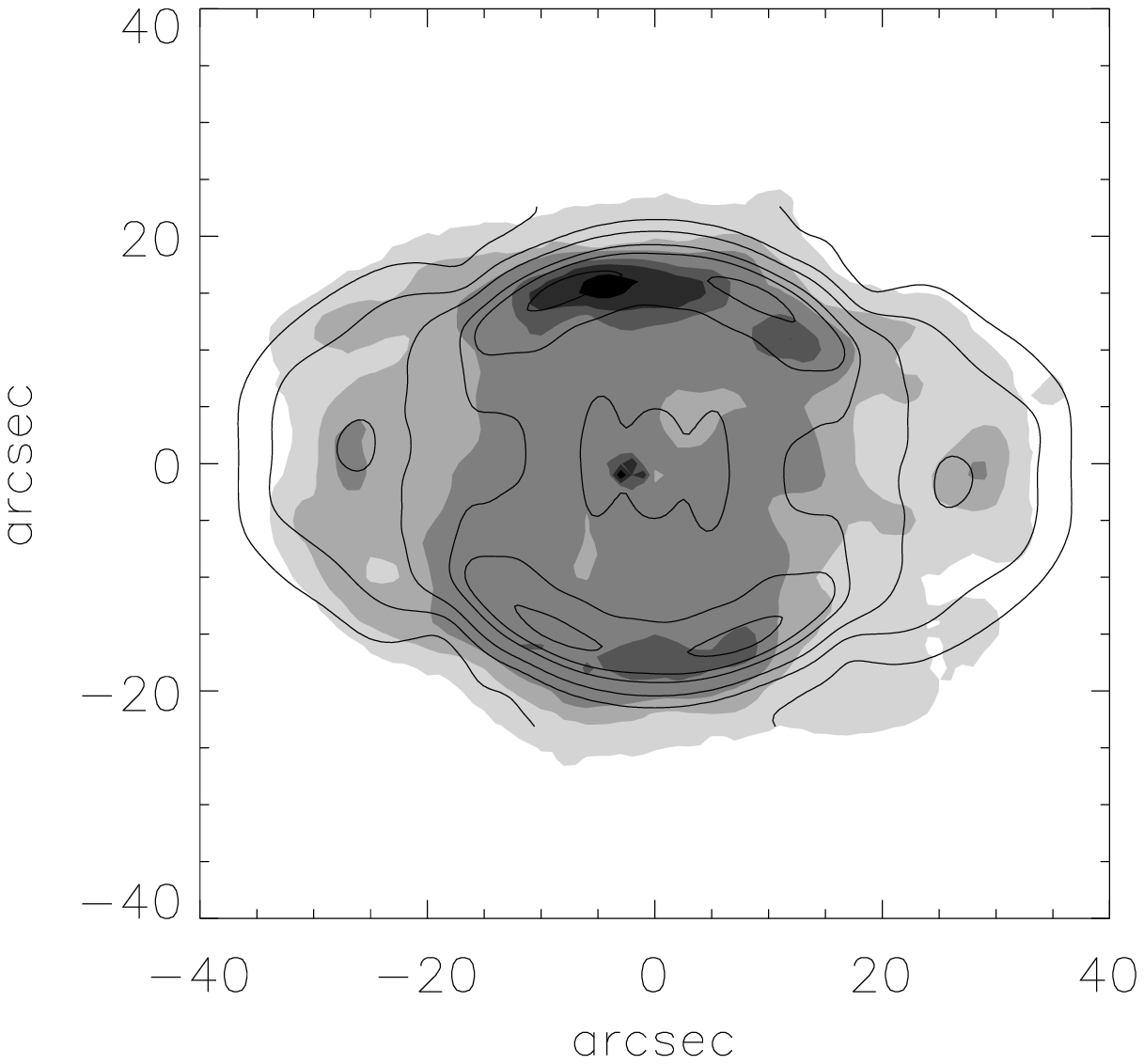}
\caption{ Image comparing the observed $H\alpha$ narrow band image
  with the contours of the equivalent image from the fitted
  model. \label{n2comp6781} }
\end{figure}

In Figure 8 we present the total SED obtained from the best fit model
(solid line), compared to the ISO spectra (gray dashed line),
UBVRI-JHK (for the central source only) and IRAS photometric data
(circles). Here, as mentioned before, we assumed the dust to be
perfectly mixed with the gas and adopted a mass dust to gas ratio of
0.0015, with a typical MRN size distribution and composition. The dust
to gass mass ratio is in good agreement with the value 0.0013 obtained
\cite{SS99}.


\begin{figure}[!h]
\includegraphics[width=\columnwidth]{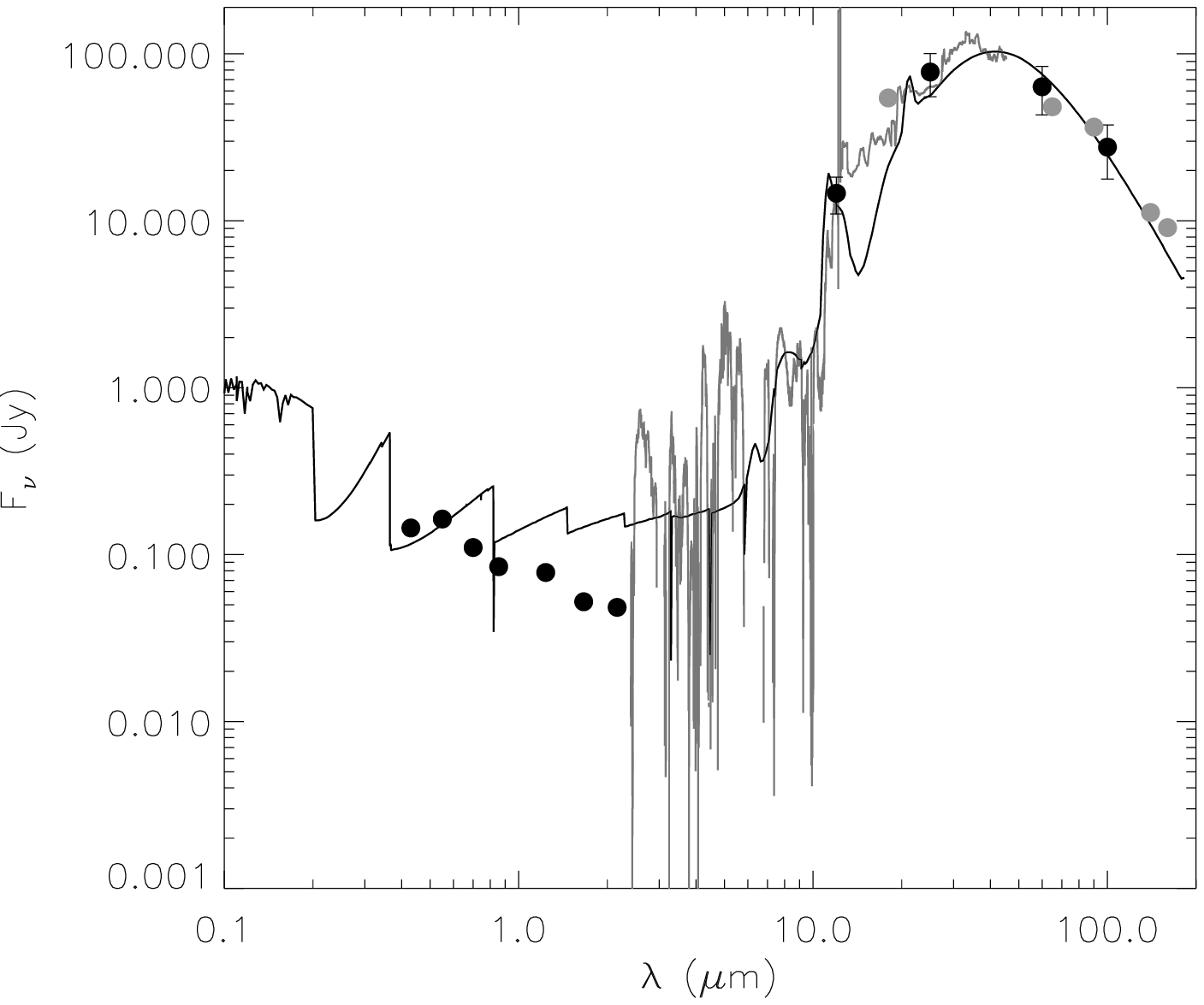}
\caption{ Model SED (solid line) for NGC~40 compared to ISO spectra
  (gray dashed line) and UBVRI, JHK (both for central source only) and
  IRAS photometric data obtained from SIMBAD. }
\end{figure}

\section{Position-Velocity Diagram}

One of the advantages of combining hydrodynamical simulations with
photionization calculation is the possibility of obtaining
self-consistent PV-diagrams. The synthetic PV-diagrams for NGC~40 can
be directly compared to those obtained observationally by
\cite{SCBTZ00}.

From the hydrodynamical simulations we obtained the velocity field
({\bf u}) and from the photoionization code MOCASSIN-3D we obtained
the emissivity for the main nebular spectral lines
($\epsilon^{\lambda}$).  The line profiles ($I_{v}$) are obtained as
\citep{falceta06}:

\begin{equation}
  \medskip
  I_{v} = \sum_{i=1}^{i_{max}} \frac{\epsilon_i^{\lambda}}{(2 \pi \sigma^{2})^{1/2}} \exp 
  \left[ -\frac{\left( v-u_{\rm LOS}^{i}\right) ^{2}}{2\sigma^{2}}\right], 
  \medskip
\end{equation} 

\noindent
where the indices $i$ represent each cell of the cube intercepted by
the LOS, $u_{\rm LOS}$ is the projected velocity along the LOS,
$\epsilon_i^{\lambda}$ is the emission of a given line $\lambda$ at
the $i$-th cell and $\sigma$ is the thermal Doppler broadening.

The result of Equation 3 is a three dimensional variable
P$\times$P$\times$V, that may be converted into a standard PV-diagram
once a specific direction (the position angle) is chosen, representing
the observational long-slit. Detailed observed PV-diagrams for NGC~40
were presented by \cite{SCBTZ00}. In this section we compare their
data with the synthetic line profiles obtained from the simulations.
Both observational (red circles) and synthetic (greyscale) data for
$[$N~{\sc ii}$]$~6584 are shown in Figure 9. The synthetic spectral
line profiles were obtained for different inclination angles with
respect to the LOS, ranging from $\theta = 0^{\circ}$ (left) to
$40^{\circ}$ (right), and for two position angles for the slit: $top$
- along the minor axis of symmetry, and $bottom$ - along the major
axis of symmetry of the nebula. Visually it is possible estimate a
best match for $\theta \sim 20^{\circ}$.

\begin{figure*}[!ht]
   \centering
   \includegraphics[width=3cm]{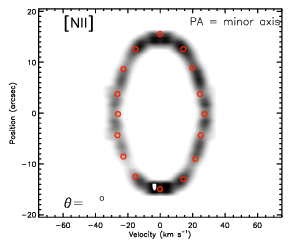}
   \hspace{0.1cm}
   \includegraphics[width=3cm]{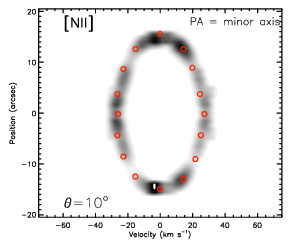}
   \hspace{0.1cm}
   \includegraphics[width=3cm]{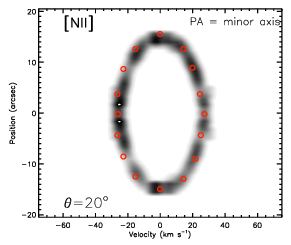}
   \hspace{0.1cm}
   \includegraphics[width=3cm]{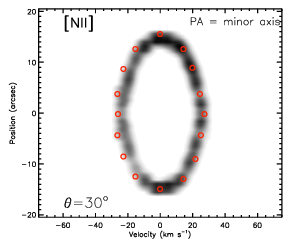}
   \hspace{0.1cm}
   \includegraphics[width=3cm]{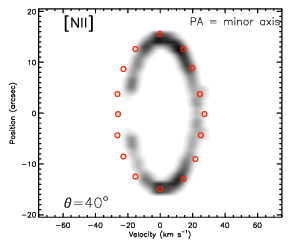}\\[10pt]
   \includegraphics[width=3cm]{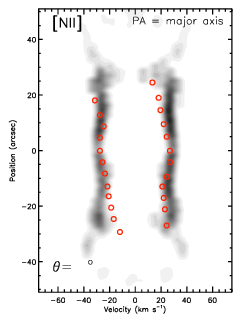}
   \hspace{0.1cm}
   \includegraphics[width=3cm]{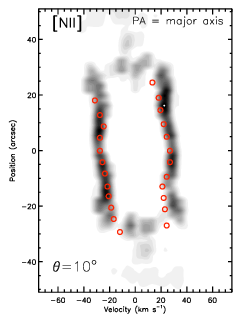}
   \hspace{0.1cm}
   \includegraphics[width=3cm]{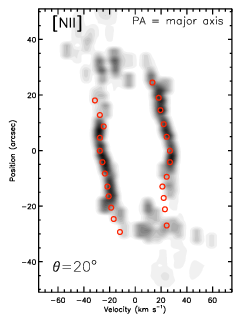}
   \hspace{0.1cm}
   \includegraphics[width=3cm]{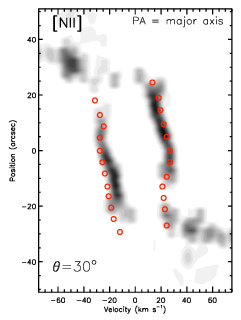}
   \hspace{0.1cm}
   \includegraphics[width=3cm]{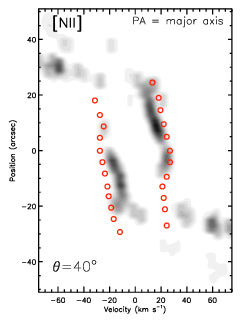}
   \caption{Position-Velocity diagrams obtained from the emission maps
     of [NII] combined with the velocity cubes from the hydrodynamical
     simulation, for different inclination angles of the PN axis of
     symmetry and the LOS (0, 10$^{\circ}$, 20$^{\circ}$,
     30$^{\circ}$, 40$^{\circ}$, respectively). Top row represents the
     position angle along the minor axis of the PN projected at the
     plane of the sky, while the bottom row is for the major
     axis. Circles represent observed values extracted from figs. 3
     and 4 of \cite{SCBTZ00}.}
\end{figure*}

\begin{figure*}[!ht]
   \centering
   \includegraphics[width=3.8cm]{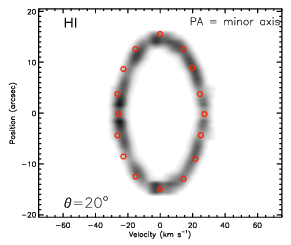}
   \hspace{0.1cm}
   \includegraphics[width=3.8cm]{figures/plot_nii_pa_minor_ang20.eps}
   \hspace{0.1cm}
   \includegraphics[width=3.8cm]{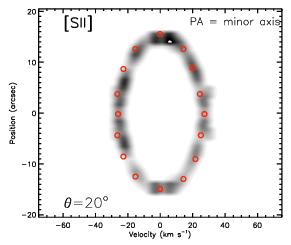}
   \hspace{0.1cm}
   \includegraphics[width=3.8cm]{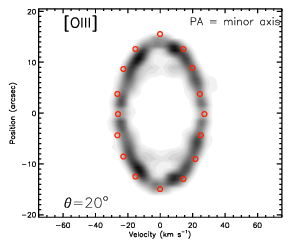}\\[10pt]
   \includegraphics[width=3.8cm]{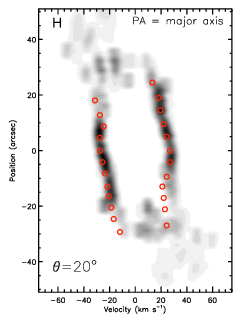}
   \hspace{0.1cm}
   \includegraphics[width=3.8cm]{figures/plot_nii_pa_major_ang20.eps}
   \hspace{0.1cm}
   \includegraphics[width=3.8cm]{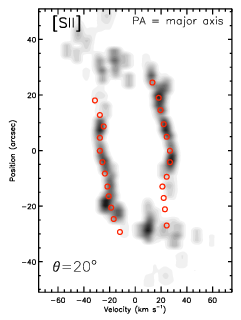}
   \hspace{0.1cm}
   \includegraphics[width=3.8cm]{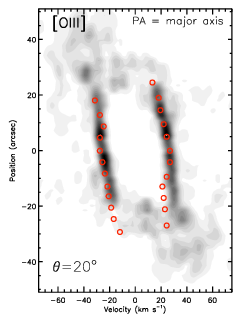}
   \caption{Position-Velocity diagrams obtained from the emission maps
     of H, [NII], [SII] and [OIII] combined with the velocity cubes
     from the hydrodynamical simulation, for an inclination angle of
     20$^{\circ}$. Top row represents the position angle along the
     minor axis of the PN projected at the plane of the sky, while the
     bottom row is for the major axis. Circles represent observed
     values for [NII] extracted from figs. 3 and 4 of \cite{SCBTZ00}.}
\end{figure*}

In Figure 10 we present the synthetic PV-diagrams obtained from the
emission maps of H, [NII], [SII] and [OIII], respectively. The circles
represent the observed data for $[$N~{\sc ii}$]$~6584, as
comparison. The plots obtained for H, [NII], [SII] are very
similar. The knots are visible as the darker spots at the edges of the
position angle (PA$\sim \pm30^{\circ}$), with a expanding velocity
$|v| \sim 10$km s$^{-1}$. These clumps are not detected in the [OIII]
maps, confirming its low ionization. On the other hand, the [OIII]
PV-diagram obtained along the minor axis of symmetry (top), show
emission at positions closer to the central source, compared to H,
[NII], [SII]. These general results corroborate with the observations
of \cite{SCBTZ00}, and shows that both the density
distribution and the velocity field obtained from the simulations
agree with the observational data of NGC~40.

\section{Discussion and conclusions}

\begin{table*}[!ht]
\begin{center}
  \caption{Comparison of literature abundance determinations for NGC\,40.}
\begin{tabular}{lccccc}
\hline
work  &     He/H  &    O/H   &    N/H   &    S/H   &   C/H    \\
\hline
Perinotto & 4.73e-2 &   7.24e-4 &    8.86e-5 &    2.80e-6 &   -       \\
Liu       &  1.2e-1 &   4.9e-4  &    8.5e-5  &    2.6e-6  &   6.9e-4  \\
Clegg     & $>0.044$&   8.4e-4  &    2.4e-4  &    3.9e-6  &   1.0e-3  \\
Pottasch  & $>0.046$&   5.3e-4  &    1.3e-4  &    5.6e-6  &   1.9e-3  \\
LGMR10    & 1.2e-1  &   4.1e-5  &    -       &    -       &   -       \\
This work & 0.9e-1  &   1.9e-4  &    6.8e-5  &    6.8e-6  &   6.5e-4  \\
\hline
\end{tabular}
\end{center}
\end{table*}

We have obtained a self consistent three dimensional model for the PN
NGC~40. Unlike previous works we have defined the three dimensional
density structure by using a 2.5D hydrodynamic simulation as described
in Sec. 3. With the HD density structure we then proceeded to the
usual 3D photoionization modeling procedure performed and described in
detail in previous papers of this series.

Though the hydrodynamical simulations do include the effects of
radiative cooling, we do not treat the effects of heating and
ionization from the central source. However, ionization fronts are
known to significantly change the propagation of shock fronts under
certain conditions (see Henney et al. 2005 for review). From Henney et
al. 2005, the role of photoionization advection is related to the
advection parameter $\lambda_{\rm ad} = \frac{\xi_{\rm ad}M}{\tau -
  \xi_{\rm ad}M}$, defined as the ratio between the fluxes of atoms
and photons at the ionization front, being $M$ the mach number, $\xi
\sim 12$ for typical central star parameters and sound speed at the
ionization front, and $\tau \sim 10^{-18} n z$, with $z$ being the
characteristic Stromgren distance. For NGC40, at the evolutionary
stage analyzed in this work, $z \sim 10^{17} – 10^{18}$ and $n \sim
3000$, we obtain $\tau \sim 300 - 3000$, which leads to $\lambda_{\rm
  ad} \sim 0.004 - 0.04 M$, i.e. the thickness of the ionization front
is very small. In such situation the effects of the ionization front
in the dynamical evolution of the nebula are negligible, and the
instabilities are in general quenched (Williams 2002).

The density and projected column densities obtained from the numerical
simulations agree with the observed large scale morphology of
NGC~40. The column density projections however do not show the knots
at the major axis of simmetry of the nebula as observed at low
ionization energy lines, such as [NII]6584. After using the simulated
nebula as input for the photoionization code MOCASSIN 3-D, the
resulting emissivities in this line showed good agreement with the
observed maps also in smaller morphological scales such as the
knots. This reveals the importance of combining both hydrodynamical
and photoionization models in order to study the true nature of the
nebular plasma.

\cite{MXS02} reported the detection of emission-line structures due to
the interaction of the PN with the interstellar medium (ISM)
attributed to Rayleigh-Taylor instability. Indeed, this is observed in
the simulations.

In the X-rays band, NGC~40 presents a faint and diffuse emission
distributed within a partial annulus of about 40'' diameter
\citep{MKMS05}. The morphology, temperature and luminosity inferred
from these observations indicate that the emission arises from a hot
bubble generated by shocked quasi-spherical fast wind from the central
star. The results also show no evidence for collimated jets. In the
simulations, as shown in Figure 1, the reverse shock interacts with
the continuous wind from the central source creating a low density and
high temperature medium. This region may be associated to the
observations in X-rays.

The results show good agreement with all observables chosen as
constraints to the modeling. The usual diagnostic maps (density,
temperature) as well as all emission line maps obtained by LGMR10 are
nicely reproduced. We also reproduce with a good degree of agreement
the kinematical data obtained by \cite{SCBTZ00}.

The total line fluxes as well as the total
$H\beta$ flux are reproduced within the observational uncertainties
for the fitted distance of $1150 \pm 120$. The uncertainty in the
distance was estimated based on the relative errors of observed total
line fluxes relative to model values and estimated errors of observed
and model emission line map characteristics done by visual
comparison. This result is in good agreement with the average result
and uncertainty obtained from many different literature values. The
good agreement of the size obtained with our structure and distance
value is clearly seen in Figure 5, where we overlay the model contours
of a simulated $H\alpha$+15\%[NII] image with the respective observed one
from a $H\alpha$ narrow band filter.

\begin{figure}[!ht]
\includegraphics[width=\columnwidth]{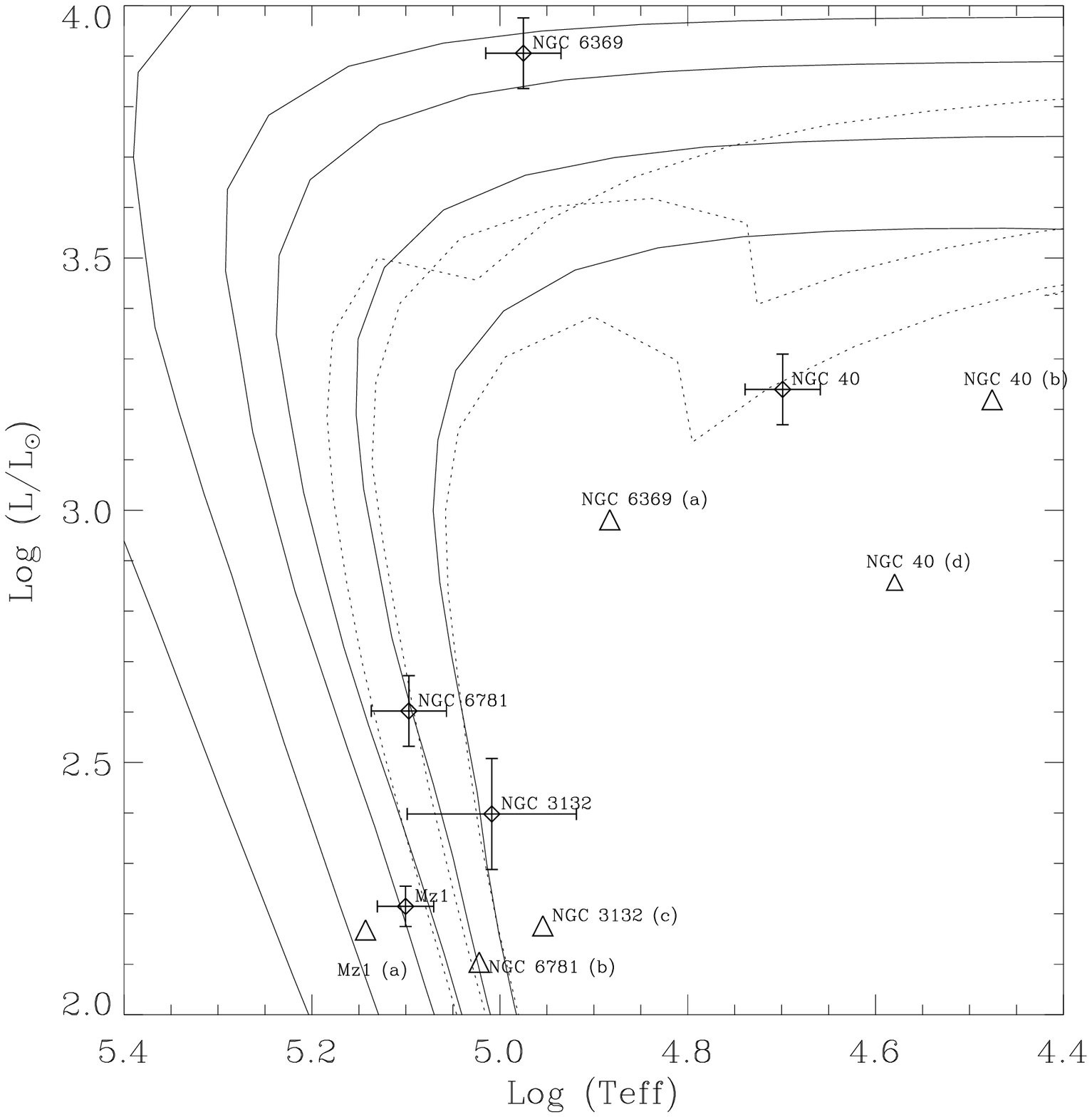}
\caption{HR diagram for NGC\,6781, NGC\,3132 (\cite{MMGV00}),
  NGC\,6369 (\cite{MSGH04}), MZ\,1 (\cite{MSGGH05}), all PNe that had
  their central star properties determined by our method. Also plotted
  are the literature values for comparison. a) \cite{SCS93}; b)
  \cite{SVMG02}; c)\cite{BDG90}d)\cite{PB10}. The evolutionary tracks are from
  \cite{VW94}; they are similar to the \cite{B95} models but take
  metallicities into account.  . \label{evolallpn} }
\end{figure} 

In Figure 11 we compare the values of effective temperature and
luminosity obtained from our model, $T_{eff}=(50 \pm 5)~kK$ and
$L=(1719 \pm 170)~L_{\odot}$, to evolutionary tracks of \cite{VW94}
and estimate an age of $(5810 \pm 600)$ yr, for a core mass of $(0.567
\pm 0.06)$M$_{\odot}$ formed from a progenitor with $(1 \pm
0.1)~M_{\odot}$. The age value is in good agreement with literature
values mentioned in Section 1.

The characteristics of the ionizing source and position in the $T_{eff}
\times L$ diagram, as shown in Figure 9, are consistent with a [WC]
star in transition to a PG1159 type object. This evolutive connection
with sequence
[WCL]$\rightarrow$[WCE]$\rightarrow$([WC]+PG1159)$\rightarrow$PG1159
was proposed by \cite{KDR98}. It is interesting to note the
discrepancy between our model results and previous ones from empirical
methods from the literature. These discrepancies are also present in
most other objects studied using our method. The lower number of
assumptions and simplifications as well as the large amount of
observational constraints used in our work indicate that the values
obtained are likely more accurate.

In \cite{MG09} the authors argue that the method of determining
distances with self-consistent modeling could be wrong if the
existence of a clumping factor is assumed. While this is theoretically
true, the use of clumping factors are not justified in the methods
developed in our work where the 3D structure used is defined down to
the cube cell resolution scale. This is especially true in the present
work since the fluctuations that appear in the density structure we
have obtained are solely due to hydrodynamical processes and no
artificial factor is needed.

In any case, we follow the exercise suggested by \cite{MG09} in their
Sec. 5.4 to determine the clumping factor. We have obtained a relation
from the \cite{VW94} model tracks in the temperature range of the
effective temperature obtained by our models for NGC 40. The relation
we obtained was $log(age)=12.025-2.543log(L/L_{\odot})$. It is
important to point out that it is not clear from \cite{MG09} how
exactly their relation was obtained. We selected points from the
evolutionary tracks of the He-burning models in the temperature range
of $4.6<log(T_{eff})<4.8$ and fitted a straight line to the region
close to the luminosity range needed.

Using the luminosity and dynamical age obtained from our model fits we
find the intersection to the relation above and obtain a value of
$k~1$. Given that this method is very crude and indirect and errors
are difficult to estimate it is impossible to obtain a reliable
confidence interval. If we use the lowest literature value for the
dynamical age, 3500 years \citep{KDR98}, we obtain $k~1.2$. Given that
the relative error quoted for the age is about 15\%, we consider this
to be consistent with a clumping factor of 1, within the errors.

In line with all the observational constraints used, although very
crude and indirect, the calculations above give some independent
evidence that there is no need for the artificial clumping factor,
indicating also the the distance determined for NGC~40 is reliable.

We have also included dust in our model, even though we have not
explored the added parameter space to fully constrain the
possibilities. However, with the assumption that the dust is mixed
with the gas and adopting a dust to gas ratio by mass of 0.0015 as
well as a typical MRN composition, we were able to reproduce the main
characteristics of the observed infrared features as can be seen in
Figure 6. A discrepancy is seen in the interval of $10\micron$ to
$20\micron$ with the continuum showing lower values than observed. The
difference could be due to a denser inner region (and thus hotter),
such as a disk, which we have not included.

In Table 3 we compare the abundances form other works to the values
obtained from our model fit. The first thing to note is the lack of
agreement between empirical literature determinations as well as
reliable confidence intervals. This is somewhat troubling since all
the empirical methods are quite similar. The major differences are
which portion of the nebula was considered and which ionization
correction factor was used, which indicates that much more work needs
to be done in these areas if a reliable empirical method is to be
developed. Even so we can say that the results are roughly correct in
terms of order of magnitude. It is important to note however, that our
method does not rely on artificial correction factors such as filling
factors or ionization correction factors, leading to results less
dependent on assumptions.

With all the considerations above we believe that our model ties in
nicely many pieces of the NGC~40 puzzle. However much still needs to
be done to definitely eliminate discrepancies, especially  in abundance
determinations as well defining the evolutionary state of these
objects, and the good agreement of many different facets of the
modeling process as done in this work suggests a possible way to
pursue this.

\begin{acknowledgements}
  H. Monteiro thanks CNPq financial support (No.\ 470135/2010-7) and
  D.F.G. thank the financial support of the Brazilian agency FAPESP
  (No.\ 2009/10102-0)
 
\end{acknowledgements}


\begin{thebibliography}

\bibitem[Acker et al.(1992)]{AO92} Acker, A., Ochsenbein, F., Stenholm, B., 
Tylenda, R., Marcout, J., Schohn, C. 1992, Strabourg-ESO Catalogue of 
Galactic PNe.

\bibitem[Akashi \& Soker(2008)]{AS08} Akashi, M., \& Soker, N.\ 2008,
  \mnras, 391, 1063

\bibitem[Balick(1987)]{B87} Balick, B.\ 1987, \aj, 94, 671 

\bibitem[Balick et al.(1992)]{BGFJ92} Balick, B., Gonzalez, 
G., Frank, A., \& Jacoby, G.\ 1992, \apj, 392, 582

\bibitem[Baessgen et al.(1990)]{BDG90} Baessgen, M., Diesch, C., \&
  Grewing, M.\ 1990, \aap, 237, 201

\bibitem[Bianchi(1992)]{B92} Bianchi, L.\ 1992, \aap, 253, 447 

\bibitem[Bl\"{o}cker(1995)]{B95} Bl\"{o}cker, T.\ 1995, \aap, 299, 755 

\bibitem[Carrasco et al.(1983)]{CSC83} Carrasco, L., Serrano, A., \&
  Costero, R.\ 1983, Revista Mexicana de Astronomia y Astrofisica, 8, 187

\bibitem[Chu et al.(1987)]{CJA87} Chu, Y.-H., Jacoby, G.~H., \&
  Arendt, R.\ 1987, \apjs, 64, 529

\bibitem[Clegg et al.(1983)]{CSPT83} Clegg, R.~E.~S., Seaton, 
M.~J., Peimbert, M., \& Torres-Peimbert, S.\ 1983, \mnras, 205, 417 


\bibitem[Cohen et al.(2002)]{CBL02} Cohen, M., Barlow, M.~J., Liu,
  X.-W., \& Jones, A.~F.\ 2002, \mnras, 332, 879

\bibitem[Ercolano et al.(2003)]{EBSL03} Ercolano, B., Barlow, 
M.~J., Storey, P.~J., \& Liu, X.-W.\ 2003, \mnras, 340, 1136 

\bibitem[Ercolano et al.(2005)]{EBS05} Ercolano, B., Barlow, 
M.~J., \& Storey, P.~J.\ 2005, \mnras, 362, 1038 

\bibitem[Falceta-Gon\c calves, Abraham \&
  Jatenco-Pereira(2006)]{falceta06} Falceta-Gon\c calves D., Abraham
  Z. \& Jatenco-Pereira, V. 2006, \mnras, 383, 258

\bibitem[Falceta-Gon\c calves et al.(2010a)]{falceta10a} Falceta-Gon\c calves 
D., de Gouveia Dal Pino E. M., Gallagher J. S. \& Lazarian A. 2010a, \apj, 708, L57

\bibitem[Falceta-Gon\c calves et al.(2010b)]{falceta10b} Falceta-Gon\c calves 
D., Caproni, A., Abraham, Z., de Gouveia Dal Pino E. M. \& Duarte, D. M. 
2010b, \apj, 713, L74

\bibitem[Falceta-Gon\c calves, Lazarian \& Houde(2010)]{falceta10c}
  Falceta-Gon\c calves D., Lazarian A. \& Houde M. 2010, \apj, 713,
  1376
 
\bibitem[Garc{\'{\i}}a-Segura et al.(2005)]{GLF05}
  Garc{\'{\i}}a-Segura, G., L{\'o}pez, J.~A., \& Franco, J.\ 2005,
  \apj, 618, 919

\bibitem[Hony et al.(2001)]{HWT01} Hony, S., Waters, L.~B.~F.~M., \&
  Tielens, A.~G.~G.~M.\ 2001, \aap, 378, L41

\bibitem[Icke et. al.(1989)]{IPB89} Icke, V., Preston, H. L. \& Balick,
  B. 1989, \aj, 97, 462

\bibitem[Icke et al.(1992)]{IBF92} Icke, V., Balick, B., \& Frank, A.\
  1992, \aap, 253, 224

\bibitem[Koesterke et al.(1998)]{KDR98} Koesterke, L., Dreizler, S.,
  \& Rauch, T.\ 1998, \aap, 330, 1041


\bibitem[Kwok(2008)]{K08} Kwok, S.\ 2008, IAU Symposium, 
252, 197

\bibitem[Leal-Ferreira et al.(2010)]{LGMR10} Leal-Ferreira, 
M.~L., Gon{\c c}alves, D.~R., Monteiro, H., 
\& Richards, J.~W.\ 2010, \mnras, 1759

\bibitem[Liu et al.(2004)]{LLLB04} Liu, Y., Liu, X.-W., Luo, 
S.-G., \& Barlow, M.~J.\ 2004, \mnras, 353, 1231 

\bibitem[Londrillo \& Del Zanna(2000)]{londrillo00} Londrillo, P. \&
  Del Zanna, L. 2000, \apj, 530, 508

\bibitem[Martin et al.(2002)]{MXS02} Martin, J.,
  Xilouris, K., \& Soker, N.\ 2002, \aap, 391, 689

\bibitem[Meaburn et al.(1996)]{MLBM96} Meaburn, J., Lopez, J. A.,
  Bryce, M., \& Mellema, G. 1996, A\&A, 307, 579

\bibitem[Monteiro, Morisset, Gruenwald, \&
Viegas(2000)]{MMGV00} Monteiro, H., Morisset, C.,
Gruenwald, R., \& Viegas, S.~M.\ 2000, \apj, 537, 853

\bibitem[Monteiro, Schwarz, Gruenwald, \& Heathcote(2004)]{MSGH04}
  Monteiro, H., Schwarz, H.E., Gruenwald, R., \& Heathcote,S.R. 2004,
  \apj, 609,194

\bibitem[Monteiro, Schwarz, Gruenwald, Guenthner, \&
  Heathcote(2005)]{MSGGH05} Monteiro, H., Schwarz, H.E., Gruenwald,
  R., Guenthner, K., \& Heathcote,S.R. 2005, \apj, 620,321

\bibitem[Schwarz \& Monteiro(2006)]{SM06} Schwarz, H.~E., \& Monteiro,
  H.\ 2006, \apj, 648, 430

\bibitem[Montez et al.(2005)]{MKMS05} Montez, R., Jr., Kastner, J.~H.,
  De Marco, O., \& Soker, N.\ 2005, \apj, 635, 381

\bibitem[Morisset \& Georgiev(2009)]{MG09} Morisset, C., \& Georgiev,
  L.\ 2009, \aap, 507, 1517

\bibitem[Phillips \& Marquez-Lugo(2011)]{PM11}
  Phillips, J.~P., \& Marquez-Lugo, R.~A.\ 2011, arXiv:1102.0526

\bibitem[Pottasch \& Bernard-Salas(2010)]{PB10} Pottasch, S.~R., \&
  Bernard-Salas, J.\ 2010, \aap, 517, A95

\bibitem[Sabbadin et al.(2000)]{SCBTZ00} Sabbadin, F., Cappellaro, E.,
  Benetti, S., Turatto, M., \& Zanin, C.\ 2000, \aap, 355, 688

\bibitem[Stanghellini, Corradi, Schwarz(1993)]{SCS93} Stanghellini,
  L., Corradi, R.L.M., Schwarz, H.E. 1993, \aap, 279,521

\bibitem[Stanghellini et al.(2002)]{SVMG02} Stanghellini, L.,
  Villaver, E., Manchado, A., \& Guerrero, M.~A.\ 2002, \apj, 576, 285

\bibitem[Stasi{\'n}ska \& Szczerba(1999)]{SS99} Stasi{\'n}ska, G., \&
  Szczerba, R.\ 1999, \aap, 352, 297

\bibitem[Vassiliadis, \& Wood(1994)]{VW94} Vassiliadis, E., Wood, P.R.
  1994, \apjs, 92, 125

\bibitem[Werner \& Herwig(2006)]{WH06} Werner, K., \&
  Herwig, F.\ 2006, \pasp, 118, 183

\end{thebibliography}
\end{document}